\documentclass[aps,prl,groupedaddress,showpacs]{revtex4}
\usepackage{graphicx}
\usepackage{dcolumn}
\usepackage{bm}
\usepackage{color}
\begin{document}

\title{On Bose-Einstein condensation in quasi-2D systems with
applications to high T$_c$ superconductivity II.}

\author{C. Villarreal$^1$ and M. de Llano$^2$  }
\date{\today}

\affiliation{$^1$ Instituto de F\'{\i}sica, Universidad Nacional Aut\'onoma de
M\'exico, Circuito de la
Investigaci\'on Cient\'ifica, Ciudad Universitaria,
CP 04510 Distrito Federal, M\'exico \\
$^2$ Instituto de Investigaciones en Materiales,
Universidad Nacional Aut\'onoma de M\'exico, Circuito de la
Investigaci\'on Cient\'ifica, Ciudad Universitaria,
CP 04510 Distrito Federal, M\'exico.
}

\pacs{ 05.30.Fk, 74.20.-z, 74.72.-h} \keywords{BEC,BCS, high-$T_c$ superconductivity}

\begin{abstract}
We calculate the number and energy densities of a quasi-2D Bose-Einstein gas
constrained within a thin region of infinite extent but of finite width $%
\delta $. The BEC critical transition temperature then becomes an explicit
function of $\delta $. We use this result to construct a model of high-$T_{c}
$ superconductivity in cuprates with a periodic layered atomic structure.
The predicted behavior of the BEC $T_{c}$ agrees with recent experimental
findings in severely underdoped cuprates.
\end{abstract}

\keywords{High-$T_{c}$  superconductivity; BEC condensates; BCS theory.}

\maketitle

\section{Introduction}

Since the discovery of high-$T_{c}$ superconductivity (HTSC) in cuprates by
Bednorz and M\"{u}ller in 1986 many studies to explain the phenomenon have
been reported. Its physical origin is not yet clear. Recent measurements 
\cite{yang08}\ of photoelectron emission spectral intensities from HTSCs
suggest that bound electron Cooper pairs (CPs) form already at temperatures
higher than the critical $T_{c}$. This finding is consistent with several
theoretical efforts \cite{dellano98,FG}\ proposing that HTSC
originates from a 2D Bose-Einstein condensate (BEC) of CPs pre-existing
above $T_{c}$ and formed through a BCS-like phonon mechanism. The 2D
character of the phase transition is associated with the layered structure
of cuprates, which in the case of $YBa_{2}Cu_{3}O_{7-y}$ (YBCO) consists of
a succession of parallel layers perpendicular to the vertical c-axis with a
unit cell of height $\sim 12\mathring{A}$, and the chemical composition $%
CuO-BaO-CuO_{2}-Y-CuO_{2}-BaO-CuO$. It is generally agreed that the $CuO_{2}$
planes, which in the case of YBCO are equidistant from the central $Y$ atom
with separation $\simeq 1.5\mathring{A}$, are mainly responsible for the
superconductivity in cuprates. Contour plots of the charge distribution
derived from energy-band-structure calculations for YBCO reveal \cite%
{krakauer88} that the SC charge carriers are mainly concentrated within a
slab of width $\delta \simeq 2.15\mathring{A}$ about the $CuO_{2}$ plane.\\

BCS-like theories \cite{dellano98,FG} contemplate a Hamiltonian $H_{BCS}$
containing the kinetic energies of electrons and holes, and a pairing
interaction arising from phonon-exchange attractions that overwhelm and
Coulombic repulsion. As a consequence, bound CPs of electrons or holes with
antiparallel spins and charge $\pm 2e$ arise with an energy-momentum
relation linear at leading order, rather than quadratic, namely $\mathcal{E}%
_{K}\simeq \mathcal{E}_{0}+c_{1}\hbar K$ with $c_{1}=2v_{F}/\pi $ in 2D with 
$v_{F}$ the Fermi speed, while $K$ is the CP center-of-mass momentum
wavenumber, and $\mathcal{E}_{0}$ the familiar weak-coupling energy $%
\mathcal{E}_{0}=-2\hbar \omega _{D}\exp [-2/v_{0}N_{0}]$ for $K=0$ CPs,
where $\omega _{D}$ is the Debye frequency, $v_{0}$ the positive BCS
electron-phonon coupling constant, and $N_{0}$ the electron density of
states for one spin at the Fermi level. The linear dispersion relation is
induced \cite{dellano07}\ by the Fermi sea medium so that CPs propagate like
free massless composite particles in the Fermi sea (whereas \textit{in vacuo}
they would do so quadratically as $\hbar ^{2}K^{2}/4m^{\ast }$ if $m^{\ast }$
is the effective electron mass). Their Bose statistical nature allows them
to undergo BEC.

\section{Formalism}

We study a model of HTSC in cuprate materials comprising a quasi-2D BEC of
excited CPs of energy $\mathcal{E}_{K}\simeq \mathcal{E}_{0}+c_{1}\hbar K$
constrained to propagate within quasi-2D layers of infinite extent in the $%
a_{1}$, $a_{2}$ directions, but of finite width $a_{3}=\delta $ in the
perpendicular direction. We assume that the CP field satisfies periodic
boundary conditions (BC) along the $a_{1}$, $a_{2}$ and $a_{3}$ directions,
although the more restrictive Dirichlet or Neumann BCs may be
straightforwardly implemented. The average number density and the energy per
unit volume of the CP field are given by 
\begin{equation}
n(T)=\frac{1}{V}\int dK\frac{g(K)}{e^{\beta (\varepsilon _{K}-\mu )}-1}\text{
\ \ and\ \ \ }u(T)=\frac{1}{V}\int dK\frac{g(K)\varepsilon _{K}}{e^{\beta
(\varepsilon _{K}-\mu )}-1}  \label{number1}
\end{equation}%
where $g(K)$ is the \textit{exact} eigenmode distribution of the field
defined by $g(K)=\sum_{\{\mathbf{n\}}}\delta (K-K_{\mathbf{n}})$ with $K_{%
\mathbf{n}}^{2}=(2n_{1}\pi /a_{1})^{2}+(2n_{2}\pi /a_{2})^{2}+(2n_{3}\pi
/a_{3})^{2}$.  By introducing Poisson's summation formula it may be shown  
that $g(K)$ can be expressed in the more tractable way 
\begin{equation}
g(K)=\frac{V}{2\pi ^{2}}K\sum_{m_{1},m_{2},m_{3}}\frac{\sin \left[ K\left(
(a_{1}m_{1})^{2}+(a_{2}m_{2})^{2}+(a_{3}m_{3})^{2}\right) ^{1/2}\right] }{%
[(a_{1}m_{1})^{2}+(a_{2}m_{2})^{2}+(a_{3}m_{3})^{2}]^{1/2}}.  \label{g2}
\end{equation}%
An alternative derivation based on properties of Bessel functions  has been formerly 
employed to study the Casimir energy-momentum tensor in rectangular cavities, both at zero and
finite temperatures \cite{hjv}. We then evaluate (\ref{number1}) by
introducing the CP mode distribution (\ref{g2}) and the CP excitation energy 
$\varepsilon _{K}\equiv \mathcal{E}_{K}-\mathcal{E}_{0}\simeq c_{1}\hbar K$.
The Bose-Einstein denominators in the integrals render a rapid convergence
so that they may safely be extended to infinity. The integrals can be
computed by expanding the integrand in powers of $ze^{-x}$ with the result 
\begin{equation}
n(T)=n_{0}(T)+\frac{(k_{B}T)^{3}}{\pi ^{2}\hbar ^{2}c^{3}}%
\sum_{m_{1},m_{2},m_{3}}\sum_{m=1}^{\infty }\frac{m\ z^{m}}{\Big(%
m^{2}+\alpha _{m_{1},m_{2},m_{3}}^{2}\Big)^{2}}  \label{number3}
\end{equation}%
with $\alpha _{m1,m2,m3}^{2}=(k_{B}T/\hbar
c_{1})^{2}[(m_{1}a_{1})^{2}+(m_{2}a_{2})^{2}+(m_{3}a_{3}]^{2}$ while the
energy density is 
\begin{equation}
u(T)=\frac{(k_{B}T)^{4}}{\pi ^{2}\hbar ^{2}c_{1}^{3}}%
\sum_{m_{1},m_{2},m_{3}}\sum_{m=1}^{\infty }\frac{\Big(3m^{2}-\alpha
_{m_{1},m_{2},m_{3}}^{2}\Big)z^{m}}{\Big(m^{2}+\alpha
_{m_{1},m_{2},m_{3}}^{2}\Big)^{3}}.  \label{energy3}
\end{equation}%
It is easily checked that the usual thermodynamic limit is attained by
considering the terms with $m_{1}=m_{2}=m_{3}=0$ in the former expressions.
On the other hand, for CPs constrained to move within a layer of finite
width $\delta \ll a_{1},a_{2}$ but unconstrained along the infinite $a_{1}$, 
$a_{2}$ directions we must set $m_{1}=m_{2}=0$ in(\ref{number3}) and (\ref%
{energy3}). In that case, we introduce the dimensionless thickness variable $%
\eta \equiv k_{B}T\delta /\hbar c_{1}$ and the remaining summations over $%
m_{3}$ may be performed analytically to give%
\begin{equation}
n(T)=n_{0}(T)+\frac{(k_{B}T)^{3}}{\pi ^{2}\hbar ^{3}c^{3}}\Psi _{3}(z,\eta )%
\text{ \ \ and\ \ \ }u(T)=3\frac{(k_{B}T)^{4}}{\pi ^{2}\hbar ^{3}c^{3}}\Phi
_{4}(z,\eta )  \label{number4}
\end{equation}%
where $\Psi _{s}(z,\eta )\equiv \sum_{m=1}^{\infty }\frac{z^{m}}{m^{s}}%
f_{m}(\eta )\ $and $\Phi _{s}(z,\eta )\equiv \sum_{m=1}^{\infty }\frac{z^{m}%
}{m^{s}}g_{m}(\eta )$ with $f_{m}(\eta )=\frac{1}{2}\left[ h_{m}(\eta )+%
\frac{m\pi }{\eta }\coth \left( \frac{m\pi }{\eta }\right) \right] $, $%
h_{m}(\eta )=\left( \frac{m\pi }{\eta }\right) ^{2}\sinh ^{-2}\left( \frac{%
m\pi }{\eta }\right) $, and $g_{m}(\eta )=\frac{1}{3}\Big[\Big(h_{m}(\eta
)+\left( \frac{m\pi }{\eta }\right) \coth \left( \frac{m\pi }{\eta }\right)
(1+h_{m}(\eta ))\Big].$

\section{Thick- and thin-layer limits }

The thick-layer limit given by $\eta \gg 1$ represents unconstrained
propagation of CPs throughout the entire volume of the material as in
conventional 3D superconductivity. In this limit $h_{m}(\eta )\rightarrow 1$%
, $f_{m}(\eta )\rightarrow 1$, $g_{m}(\eta )\rightarrow 1$ and we recover
well-known expressions describing 3D BECs \cite{dellano98,FG}: 
\begin{equation}
n(T)=n_{0}(T)+\frac{(k_{B}T)^{3}}{\pi ^{2}\hbar ^{3}c^{3}}\sum_{m=1}^{\infty
}\frac{z^{3}}{m^{3}}\text{ \ \ and\ \ \ }u(T)=3\frac{(k_{B}T)^{4}}{\pi
^{2}\hbar ^{3}c^{3}}\sum_{m=1}^{\infty }\frac{z^{4}}{m^{4}}  \label{number5}
\end{equation}%
The critical temperature $T_{c}$ follows from the conditions $%
n_{0}(T_{c})\rightarrow 0$ and $z(T_{c})\rightarrow 1$ leading to $%
k_{B}T_{c}^{3D}=[\pi ^{2}\hbar ^{3}c_{1}^{3}n^{3D}/\zeta (3)]^{1/3}$ where $%
\zeta (n)$ is Riemann's $\zeta $-function. The molar heat capacity $%
C(T)=R(nk_{B})^{-1}\partial u(T)/\partial T$ (with R the gas constant) is
straightforwardly obtained from (\ref{number5}) as $C(T)=\left( \frac{%
12R\zeta (4)}{\zeta (3)}\right) \left( \frac{T}{T_{c}}\right) ^{3}$ for $%
T<T_{c}$ and $C(T)=\left( \frac{12R\zeta (4)}{\zeta (3)}\right) \left( \frac{%
T}{T_{c}}\right) ^{3}-\left( \frac{12R\zeta (3)}{\zeta (2)}\right) $ for $%
T>T_{c}$ which is consistent with measurements in conventional 3D
superconductors \cite{poole95}  since at $T=T_{c}$ the heat capacity shows a
discontinuous drop $\Delta C=6.57R$ indicative of a second-order phase
transition.

In the thin-layer limit $\eta \ll 1$ associated with HTSC we get $h_{m}(\eta
)\rightarrow 0$, $f_{m}(\eta )\simeq m\pi /2\eta $, and $g_{m}(\eta )\simeq
m\pi /3\eta $. Simple algebra leads to 
\begin{equation}
n^{2D}(T)=n_{0}^{2D}(T)+\frac{(k_{B}T)^{2}}{2\pi \hbar ^{2}c_{1}^{2}}%
\sum_{m=1}^{\infty }\frac{z^{m}}{m^{2}}\text{ \ \ and\ \ \ }u^{2D}(T)=\frac{%
(k_{B}T)^{3}}{\pi \hbar ^{2}c_{1}^{2}}\sum_{m=1}^{\infty }\frac{z^{m}}{m^{3}}%
,  \label{number2D}
\end{equation}%
where $n^{2D}\equiv n\delta $ and $u^{2D}\equiv u\delta $. The critical BEC
temperature is now given by $k_{B}T_{c}^{2D}=[2\pi \hbar
^{2}c_{1}^{2}n^{2D}/\zeta (2)]^{1/2}$ where $\zeta (2)=\pi ^{2}/6$. In this
case the molar heat capacity is $C(T)=[6R\zeta (3)/\zeta (2)]\left(
T/T_{c}\right) ^{2}$ for $T<T_{c}$ but it must be evaluated numerically for $%
T>T_{c}$. It turns out that the $C(T)$ is continuous at $T=T_{c}$ although
its derivative $\partial C/\partial T$ is discontinuous. The linear behavior 
$C(T)/T\propto T$ for $T\leq T_{c}$ is characteristic of cuprate materials 
\cite{fisher88}.

Of crucial importance in evaluating $T_{c}$ is to reliably estimate the
fraction of charge carriers that actually contribute to the supercurrent.
The charge carrier density is usually determined from measurements of London
penetration depth $\lambda _{ab}$ along the $CuO_{2}$ planes. It gives an
estimate of the supercurrent that causes partial rejection of an applied
external magnetic field in the superconductor. Within the framework of the
present model the supercurrent is due to massless-like CPs of charge $2e$
moving with the CP speed $c_{1}$, so that the surface supercurrent $\mathbf{J%
}_{s}=n^{2D}(2e)c_{1}\hat{\mathbf{k}}$ with $\hat{\mathbf{k}}\equiv \mathbf{k%
}/k$ \cite{FG}. A straightforward calculation \cite{villarreal09} shows that 
$n^{2D}=(e^{2}/32\pi c_{1}^{2}c^{2})\delta \Delta _{0}^{2}/\hbar \omega
_{D}\lambda _{ab}^{2}$ where $c$ is the speed of light and $\Delta _{0}$ is
the zero-temperature energy gap. The final expression of the BEC critical
temperature is 
\begin{equation}
T_{c}=\frac{\hbar c}{2\pi k_{B}e}\left( \frac{3\delta }{2\hbar \omega _{D}}%
\right) ^{1/2}\frac{\Delta _{0}}{\lambda _{ab}}.  \label{critical3}
\end{equation}

\section{Conclusions}

Introducing in (\ref{critical3}) the YBCO parameters tabulated in Ref.\cite%
{poole95} $\Theta _{D}=410$ K , $\Delta _{0}=14.5$ meV, and $\delta =2.15\ 
\mathring{A}$ \cite{krakauer88} we get the relation $T_{c}=16.79/\lambda
_{ab}$ ($\mu $m-K) which accurately reproduces experimental data reported by
Zuev \textit{et al.} \cite{zuev05} in measurements performed in YBCO films
with $T_{c}$s ranging from $6$ to $50K$. They conclude that, within some
noise their data fall on the same curve $\lambda _{ab}^{-2}\propto
T_{c}^{2.3\pm 0.4}$ regardless of annealing procedure, oxygen content, etc.
In an independent study, Broun \textit{et al.} \cite{broun07} found that
their samples of high-purity single-crystal YBCO followed also the rule $%
T_{c}\propto \lambda _{ab}^{-1}$. A forthcoming paper \cite{villarreal09}
gives a more detailed discussion of the model presented here applied also to
several other cuprates including the $l$-wave extension of the present
formalism valid only for $l=0$.\\

\textbf{Acknowledgments }We thank D.M. Eagles, M. Fortes, S. Fujita and M.A.
Sol\'{\i}s for fruitful discussions. MdeLl thanks UNAM-DGAPA-PAPIIT (Mexico)
IN106908 for partial support and is grateful to W.C. Stwalley for
discussions and the University of Connecticut for its hospitality while on
sabbatical leave.


\begin{thebibliography}{99}

\bibitem{yang08} Yang H B, Rameau J\ D, Johnson P\ D, Valla T, and Gu G D
2008 \textit{Nature} \textbf{456,} 77.

\bibitem{dellano07} de Llano M and Annett J F 2007 Int. J. Mod. Phys. B 
\textbf{21} 3657

\bibitem{dellano98} Casas M, Rigo A, de Llano M, Rojo O and Sol\'{\i}s M A
1998 \textit{Phys. Lett.} A \textbf{245}, 55

\bibitem{FG} Fujita S, Ito K and Godoy S 2009 \textit{Quantum Theory of
Conducting Matter: Superonductivity} (Springer-Verlag, Heidelberg)

\bibitem{krakauer88} Krakauer H, Pickett W E and Cohen R E 1998 \textit{J.
Supercond.} {\textbf{1}}, 111

\bibitem{hjv} Hacyan S, J\'auregui R and Villarreal C 1993 \textit{Phys. Rev.%
} A \textbf{47} 4204; J\'auregui R, Villarreal C and Hacyan S 2006 \textit{%
Ann. Phys.} \textbf{321}, 2156

\bibitem{villarreal09} Villarreal C and de Llano M, in preparation

\bibitem{poole95} Poole C P, Farach H A and Creswick R J 1995 \textit{%
Superconductivity} (Academic Press, London)

\bibitem{fisher88} Fisher R A, Gordon J E, and Phillips N E 1988 \textit{\
J. Supercond.} \textbf{1} 231

\bibitem{zuev05} Zuev Y, Kim M S and Lemberger T R 2005 \textit{Phys. Rev.
Lett.} \textbf{95} 137002-1

\bibitem{broun07} Broun D M, Huttema W A, Turner P J, {\"{O}}scan S, Morgan
B, Liang R, Hardy W N and Bonn D A 2007 \textit{Phys. Rev. Lett.} \textbf{99}
237003-1

\end{thebibliography}
\end{document}